# First Detection of Molecular Gas in the Giant Low Surface Brightness Galaxy Malin 1

Gaspar Galaz[1], Jorge González-López[1,2], Viviana Guzmán[1], Hugo Messias[3], Junais[4], Samuel Boissier[5], Benoît Epinat[5,6], Peter M. Weilbacher[7], Thomas Puzia[1], Evelyn J. Johnston[8], Philippe Amram[5], David Frayer[9], Matìas Blaña[10], J. Christopher Howk[11], Michelle Berg[12], Roy Bustos-Espinoza[1], Juan Carlos Muñoz-Mateos[13], Paulo Cortés[14,15], Diego García-Appadoo[16], and Katerine Joachimi[17]

[1] Instituto de Astrofísica, Pontificia Universidad Católica de Chile, Vicuña Mackenna 4860, 7820436 Macul, Santiago, Chile; ggalazl@gmail.com
[2] Las Campanas Observatory, Carnegie Institution for Science, Raúl Bitrán 1200, La Serena, Chile
[3] Joint ALMA Observatory, Alonso de Córdova 3107, Vitacura 763-0355, Santiago, Chile
[4] Instituto de Astrofísica de Canarias, Vía Láctea S/N, E-38205 La Laguna, Spain and Departamento de Astrofísica, Universidad de la Laguna, E-38206 La Laguna, Spain
[5] Aix-Marseille Univ, CNRS, CNES, LAM (Laboratoire d'Astrophysique de Marseille), Marseille, France
[6] Canada–France–Hawaii Telescope, 65-1238 Mamalahoa Highway, Kamuela, HI 96743, USA
[7] Leibniz-Institut für Astrophysik Potsdam (AIP), An der Sternwarte 16, 14482 Potsdam, Germany
[8] Instituto de Estudios Astrofísicos, Facultad de Ingeniería y Ciencias, Universidad Diego Portales, Av. Ejército Libertador 441, Santiago, Chile
[9] Green Bank Observatory, P.O. Box 2, Green Bank, WV 24944, USA
[10] Instituto de Alta Investigación, Universidad de Tarapacá, Casilla-7D, Arica, Chile
[11] Department of Physics and Astronomy, University of Notre Dame, Notre Dame, IN 46556, USA
[12] Department of Astronomy, The University of Texas at Austin, Austin, TX 78712, USA
[13] European Southern Observatory, Karl-Schwarzschild Straβe 2, 85748 Garching bei Munchen, Germany
[14] Joint ALMA Observatory, Alonso de Córdoba 3107, Vitacura, Santiago, Chile
[15] National Radio Astronomy Observatory, 520 Edgemont Road, Charlottesville, VA 22903, USA
[16] Esinel Ingenieros, Holanda 100, Providencia, Santiago, Chile
[17] Universidad Técnica Federico Santa María, Vicuña Mackenna 3939, San Joaquín, Santiago, Chile
Received 2024 September 20; revised 2024 October 9; accepted 2024 October 12; published 2024 November 1

## Abstract

After over three decades of unsuccessful attempts, we report the first detection of molecular gas emission in Malin 1, the largest spiral galaxy observed to date, and one of the most iconic giant low surface brightness galaxies. Using Atacama Large Millimeter/submillimeter Array, we detect significant $^{12}$CO ($J = 1$–0) emission in the galaxy's central region and tentatively identify CO emission across three regions on the disk. These observations allow for a better estimate of the $H_2$ mass and molecular gas mass surface density, both of which are remarkably low given the galaxy's scale. By integrating data on its H I mass, we derive a very low molecular-to-atomic gas mass ratio. Overall, our results highlight the minimal presence of molecular gas in Malin 1, contrasting sharply with its extensive, homogeneous atomic gas reservoir. For the first time, we position Malin 1 on the Kennicutt–Schmidt diagram, where it falls below the main sequence for normal spirals, consistent with previous upper limits but now with more accurate figures. These findings are crucial for constraining our understanding of star formation processes in environments characterized by extremely low molecular gas densities and for refining models of galaxy formation, thereby improving predictions concerning the formation, evolution, and distribution of these giant, elusive galaxies.

*Unified Astronomy Thesaurus concepts:* Spiral galaxies (1560); Spiral arms (1559); Molecular gas (1073); Low surface brightness galaxies (940); Detection (1911); Diffuse molecular clouds (381); Interferometers (805)

## 1. Introduction

Low surface brightness galaxies (LSBGs) exhibit surface brightness (SB) levels that are fainter than the dark night sky (M. J. Disney 1976). Current data indicate that they dominate the galaxy volume number density (J. J. Dalcanton et al. 1997; K. O'Neil & G. Bothun 2000; G. Martin et al. 2019).

Being quite difficult to detect in typical magnitude-limited optical surveys (W. J. G. de Blok & S. S. McGaugh 1997), LSBGs are characterized by low stellar and star formation densities (J. M. van der Hulst et al. 1993; J. P. E. Gerritsen & W. J. G. de Blok 1999), and their neutral hydrogen dominance makes molecular gas tracers particularly elusive (T. E. Pickering et al. 1997; K. O'Neil et al. 2000, 2023).

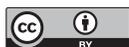

The existence of a large population of LSBGs appeared in the debate in the 1970s when, at first glance, observations from photographic plates suggested a SB plateau of 21.65 $B$ mag arcsec$^{-2}$ for disks in spiral galaxies, an idea that converged in those years to the so-called "Freeman's law" (K. C. Freeman 1970). The fact that this was proved afterward to be a bias introduced by photographic plates themselves (M. J. Disney 1976) triggered the search of LSBGs both in already available photographic plates and in new surveys (S. J. Maddox et al. 1990; C. D. Impey et al. 1996). It was in this way that the giant LSBG (gLSBG) Malin 1 was accidentally discovered, when astronomers were searching for LSBGs from photographic plates in the 1980s. Initially, it appeared as a normal spiral galaxy, surrounded by a barely visible nebulosity (G. D. Bothun et al. 1987). Subsequent H I observations revealed that the galaxy possesses an enormous H I disk, with an extent of several arcminutes in the sky (G. D. Bothun et al. 1987). However, a shocking aspect was not only the diffuse structure barely observed in the optical but also the physical size of Malin 1. The redshift measured both in the optical and in H I ($z \sim 0.082$) indicated that Malin 1 has a diameter of at least hundreds of kiloparsecs ($\sim$100–150 kpc up to a photographic isophote reaching 25 mag arcsec$^{-2}$, measured at those years).

This discovery immediately placed Malin 1 as one of the largest disk galaxies, with one of the most massive H I reservoirs ($M_{\rm H\,I} \sim (4-7) \times 10^{10} M_\odot$; C. Impey & G. Bothun 1989; T. E. Pickering et al. 1997; F. Lelli et al. 2010), and exhibiting one of the faintest disks ever observed in the optical, with a central SB of $\sim$25.5 mag arcsec$^{-2}$ in the $V$ band (C. Impey & G. Bothun 1989). A detailed spectroscopic analysis also hinted at the presence of an active galactic nucleus (AGN) in the central part of Malin 1. Initially classified as a Seyfert 1 (C. Impey & G. Bothun 1989), it was later reclassified as a low-ionization nuclear emission-line region (LINER) AGN type (A. J. Barth 2007).

Spectroscopic studies of Malin 1 revealed nebular emission lines, indicating likely stellar formation. A. J. Barth (2007) detected H$\alpha$ and other emission lines from the nucleus but did not explore stellar formation. Junais et al. (2020) also found nebular emission lines from the central regions with the Very Large Telescope (VLT). Star formation rates (SFRs) derived from H$\alpha$ fluxes suggest an early-type disk in the inner region and extended UV-like galaxies in the outer parts. They also noted high metallicity and low dust content in the inner regions.

S. Boissier et al. (2016) conducted a multiband photometric study, showing that the SB and color profiles suggest a long, quiet star formation history, with angular momentum 20 times that of the Milky Way (MW). They found an SFR of $\sim$2 $M_\odot$ yr$^{-1}$ and concluded that the star formation scenario did not require galaxy interactions.

Regarding the environment, Malin 1 interacts with two compact galaxies, Malin 1A and Malin 1B (V. P. Reshetnikov et al. 2010; G. Galaz et al. 2015; K. Saha et al. 2021), and possibly a galaxy 350 kpc away (V. P. Reshetnikov et al. 2010; G. Galaz et al. 2015). Otherwise, it is relatively isolated within 1 Mpc (R. Bustos-Espinoza et al. 2025, in preparation).

Finally, Junais et al. (2024) used VLT/MUSE to detect nebular emission lines in the disk, confirming recent star formation in several regions, as earlier suspected by other authors (G. D. Bothun et al. 1987). This work will be referenced throughout this Letter.

Current figures suggest that Malin 1 is perhaps the largest known spiral galaxy, with a disk diameter of $\sim$200 kpc and an SB fainter than 28.0 mag arcsec$^{-2}$ in the optical $B$ band, with very diffuse but thick and rich spiral arms (G. Galaz et al. 2015; S. Boissier et al. 2016; K. Saha et al. 2021).

The features described above settled the scenario for discovering the presence of molecular gas in Malin 1, which was intensively searched shortly after its discovery. However, these searches and all posterior ones, during the past 30 yr, failed (C. Impey & G. Bothun 1989; S. J. E. Radford 1992; J. Braine et al. 2000; G. Galaz et al. 2022). In summary, these figures set a quite perplexing picture: the largest spiral galaxy discovered so far, without a trace of molecular gas over its $\sim$200 kpc diameter.

Here we report the first measurement of molecular gas emission in Malin 1. Using Atacama Large Millimeter/submillimeter Array (ALMA) Band 3 and the most compact array configuration, observations in the first semester of 2024 reveal the presence of continuum emission and $^{12}$CO (1–0) line emission, with significant flux in the central part and tentative emission in three other regions in the disk.

We present observations in Section 2, and we present data analysis and some direct results in Section 3. In Section 4 we present some physical estimates and provide a discussion, and we summarize and conclude in Section 5. Throughout this Letter we assume a $\Lambda$CDM cosmology with parameters $H_0 = 70$ km s$^{-1}$ Mpc$^{-1}$, $\Omega_M = 0.3$, and $\Omega_\Lambda = 0.7$ and a heliocentric redshift for Malin 1 of $z = 0.082702$. With these parameters, Malin 1 has a distance of 370 Mpc and a plate scale of 1.57 kpc arcsec$^{-1}$.

## 2. Observations

ALMA Band 3 observations were carried on during 2024 March, using the most compact configuration C-1 of the 12 m array. The total on-source time was 5.64 hr split into two pointings separated by 39″. The first pointing was centered in the bulge of the galaxy (J2000: R.A. 12:36:59, decl. +14:19:49); the second one, in a disk region located to the northwest (J2000: R.A. 12:36:57, decl. +14:20:15). The choice of this last region was motivated by recent VLT/MUSE results, showing bright emission from this place (Junais et al. 2024). Considering the rest frequency of 115.271202 GHz for the $^{12}$CO (1–0) line, and given the heliocentric redshift of Malin 1, the observed frequency for $^{12}$CO (1–0) was tuned to 106.458562 GHz. The continuum was observed in four spectral windows, each with 1875 MHz bandwidth.

Observations were calibrated using CASA[18] version 6.5.4.9 and the ALMA pipeline version 2023.1.0.124. The imaging of the observations returns a natural weighting synthesized beam of $3\rlap{.}{''}76 \times 3\rlap{.}{''}41$ and a position angle of $-63\rlap{.}{°}9$ for the continuum image. Similar synthesized beams are obtained for the spectral cubes created for the individual spectral windows.

The first step in imaging the data was to create a multifrequency synthesis of all available spectral windows and channels and all the pointings using the mosaic option as a gridder. All images and cubes were cleaned using the auto-masking option known as "auto-multithresh." Cleaning masks were created over all the features above the $4\sigma$ significance level, where the $\sigma$ value is calculated automatically by CLEAN in the residuals using a robust rms estimate. Masks were expanded down to $2\sigma$ and cleaned down to $1\sigma$.

## 3. Data Analysis and Results

### 3.1. Detections

The final image combines the continuum and potential line emission. We are clearly detecting a source in the galaxy's bulge with a signal-to-noise ratio (S/N) above 17.

The left panel of Figure 1 shows the $^{12}$CO (1–0) line detection contours in the central part, superimposed on a composite optical image of Malin 1 from S. Boissier et al. (2016). We also show the boundaries of the region covered by our ALMA observations by a dashed line. Specifically, the line indicates where the primary beam correction reaches 0.3 (i.e., a depth of 30% relative to that reached at the phase center). The right panels show the continuum and line contours superimposed on the H$\alpha$ emission obtained recently with VLT/MUSE by Junais et al. (2024). It is worth noting the double emission feature of the H$\alpha$ emission in the central part. This has been discussed as a possible double central star cluster (E. J. Johnston et al. 2024), where significant H$\alpha$ emission is being emitted. The southern $^{12}$CO (1–0) emission in the central part coincides with the H$\alpha$ emission shown by E. J. Johnston et al. (2024), and contours at different significance levels clearly show that the detection is statistically significant at $\simeq 6\sigma$ (see Figure 1). It is apparent that the continuum seems better aligned with H$\alpha$ than the CO. This could be because of the higher S/N, giving us better precision, and the offset seen in $^{12}$CO (1–0) is just produced by low S/N. In addition, the offsets between the centroids of H$\alpha$ and $^{12}$CO (1–0) are well within the typically accepted centroid precision of ALMA observations of around 0.1–0.2 times the beam size ($\sim 4\farcs 5$), i.e., $0\farcs 45$–$0\farcs 90$.

---

[18] CASA, Common Astronomy Software Application, provided by ALMA Observatory.

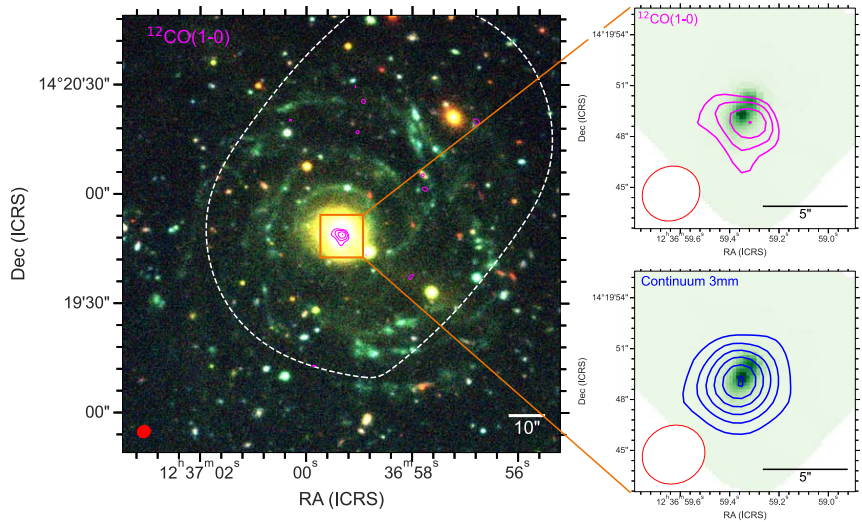

**Figure 1.** Continuum and $^{12}$CO (1–0) line emission detections with ALMA Band 3. The left panel shows a color optical image ($u$, $g$, $i$, and $z$ in blue, green, yellow, and red, respectively) from S. Boissier et al. (2016). The inset shows the central region where continuum and line emission are significant. The red ellipse in the lower left corner of each panel represents the ALMA synthesized beam size. In the main (left) panel, we show the contours of the $^{12}$CO (1–0) velocity-integrated map superimposed onto the optical image. The two panels on the right-hand side show the 3 mm continuum and $^{12}$CO (1–0) velocity-integrated flux map contours superimposed onto the H$\alpha$ emission from Junais et al. (2024). Magenta contours for the line emission over the optical and H$\alpha$ images start at $3\sigma$ and increase in steps of $1\sigma$. For the continuum emission (bottom right panel) contours start at $3\sigma$ and increase in steps of $3\sigma$.

We detect an emission line associated with the nucleus with a significance of S/N = 6.5 centered at 106.518 GHz. Figure 2 shows the $^{12}$CO (1–0) line profile, fitted to the ALMA Band 3 data for the central region described in Figure 1. It provides the flux intensity as a function of the velocity offset with respect to the optical redshift of the galaxy.

The measured deviation of $-120$ km s$^{-1}$ of the central CO emission observed in Figure 2 with the systemic velocity,[19] compared to the corresponding approximately $-20$ km s$^{-1}$ from MUSE stellar and ionized gas kinematics (E. J. Johnston et al. 2024), is compatible with the large velocity dispersion of approximately $\pm 200$ km s$^{-1}$ measured with MUSE in the central regions from both ionized gas and stars. The small offset of the CO center to the southwest with respect to the ionized gas and stellar distributions may also induce lower velocities. The relatively large velocity width in CO at the center is an additional hint that supports that this central CO detection belongs to Malin 1.

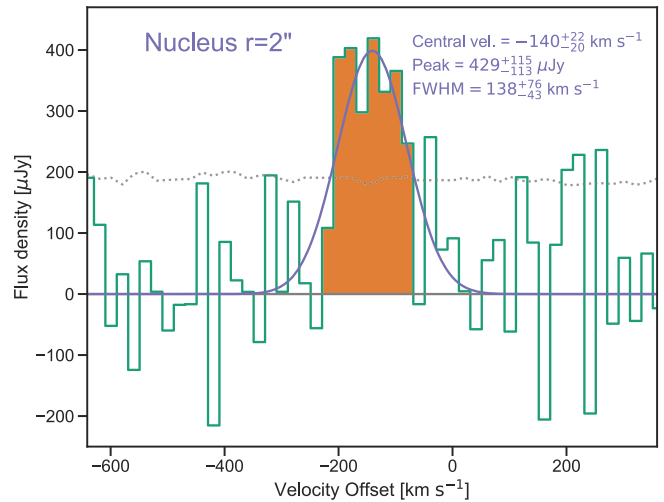

**Figure 2.** Line profile for the $^{12}$CO (1–0) central detection in Malin 1. The dotted line represents the $1\sigma$ noise level, estimated as the rms measured in the cube without primary beam correction rescaled by the square root of the ratio between the area of the aperture and the area of the synthesized beam. The line is at the $6\sigma$ significance level, which is measured in the image obtained by collapsing the orange channels and comparing the peak of the signal with the rms of the background. The line is detected at a systemic velocity of approximately $-140$ km s$^{-1}$ relative to the heliocentric redshift of the galaxy ($z = 0.082702$). The parameters of the best-fit Gaussian are shown in the upper right corner.

For the central part, the flux density for the continuum is $107 \pm 12\,\mu$Jy and the velocity-integrated flux density for the $^{12}$CO (1–0) line is $143 \pm 41$ mJy km s$^{-1}$.

To measure the integrated flux and angular extension of the emission lines, we proceeded to do a continuum subtraction both in the image plane and in the $u$-$v$ plane. We used IMCONTSUB from the CASA package to subtract the continuum emission fitted in the range of frequencies surrounding the detected line. After the continuum subtraction, a moment 0 was created, collapsing the channels with emission associated with the central line. The line emission was then fitted using IMFIT within CASA (and returned the integrated flux of $143 \pm 41$ mJy km s$^{-1}$ mentioned before). The line is also barely resolved with an FWHM major axis of $4\farcs4 \pm 1\farcs9$, a minor axis of $2\farcs6 \pm 1\farcs4$, and a position angle of $61° \pm 44°$.

---
[19] Obtained with the heliocentric redshift used in this study.

We note a high continuum emission from the center of Malin 1, which is not straightforward to explain. Some authors have noted this behavior in LINERs; see, e.g., J. J. Condon (1996), who suggested that LINERs are more likely to contain compact nuclear radio sources that cannot be obscured by dust than normal galaxies. Between 1 and 1000 GHz, a LINER spectrum is a combination of synchrotron emission, free–free emission, and thermal emission from dust. However, at 100 GHz it is very difficult to disentangle the nature of the continuum emission without more continuum frequencies (J. J. Condon 1992, 1996). Other authors have proposed alternative explanations for this behavior (R. Yan & M. R. Blanton 2012; R. Singh et al. 2013).

In order to better understand the nature of the continuum emission, we look into the NRAO VLA Sky Survey and the Faint Images of the Radio Sky at Twenty cm survey at 1.4 GHz and the Very Large Array Sky Survey (VLASS) at 3 GHz, with the aim of obtaining the radio spectral index for Malin 1, which in turn could shed light on the nature of the emission. Although the galaxy is only in VLASS at 3 GHz, it allow us to estimate the flux at 1.4 GHz assuming a spectral index. The flux at 3 GHz in VLASS is 390 $\mu$Jy beam$^{-1}$, exactly at the center of Malin 1 (at 3.0 GHz, with a $2''$ beam). The noise in the VLASS radio image is $\sim$70 $\mu$Jy beam$^{-1}$ (so at $5.5\sigma$ it is a Very Large Array detection). Assuming a spectral index of $-0.7$, this would imply 665 $\mu$Jy beam$^{-1}$ at 1.4 GHz. This is a factor of $\sim$2 lower than the value estimated from extrapolating from the 3 mm value using the M82 (starburst) radio spectral energy distribution from J. J. Condon (1992).

We then use an empirical relation for star-forming galaxies between the continuum emission at 100 GHz (3 mm) and that at 1.4 GHz by V. Heesen (2014), which in turn is related to the SFR through the formula by J. J. Condon (1992) and reformulated as SFR($M_\odot$ yr$^{-1}$) = $0.75 \times 10^{-21} \times L_{1.4\,\text{GHz}}$ (W Hz$^{-1}$). Using our value in luminosity units, we have $L_{1.4\,\text{GHz}} \sim 1.22 \times 10^{22}$ W Hz$^{-1}$. Using the V. Heesen (2014) formula, we have SFR$_{1.4\,\text{GHz}} \sim 9\,M_\odot$ yr$^{-1}$. This value is still high compared to the $0.34\,M_\odot$ yr$^{-1}$ from E. J. Johnston et al. (2024), suggesting contributions to the 3 mm continuum emission by other processes, like AGN activity, consistent with the LINER in the central part of Malin 1.

It is worth noting that the continuum emission was also subtracted from the visibilities using the task UVCONTSUB. This procedure was only done in the pointing centered in the nucleus of the galaxy since it is known that UVCONTSUB does not work properly when the continuum source is far from the phase center, as is the case for the pointing centered in the disk region. The subtraction of the continuum in the $u$-$v$ plane was done using the frequency range already used for IMCONTSUB and returned a set of visibilities without continuum. We then collapsed the visibilities to the same frequency range identified in the spectral cube and used to create the moment 0. Analysis in the $u$-$v$ plane with UVMODELFIT returned the already-mentioned values: total integrated flux of $142 \pm 21$ mJy km s$^{-1}$, with a major axis of $5'' \pm 0\farcs9$, axis ratio of $0.67 \pm 0.19$, and position angle of $34° \pm 19°$. Both estimates agree on the total flux and resolved nature of the emission.

Since the observations have marginally resolved the CO emission, we can estimate the area of a 2D Gaussian by using the following formula:[20]

$$\text{Area} = \frac{\pi \times \text{BMAJ} \times \text{BMIN}}{4 \ln 2}, \quad (1)$$

where BMAJ and BMIN correspond to the major and minor axes of the 2D Gaussian, respectively. The results from IMFIT return an area of $\approx$32 kpc$^2$, while the result from the $u$-$v$ plane analysis returns a slightly larger area of $\approx$47 kpc$^2$. Conservatively, the emission area is estimated to be within $(3–5) \times 10^7$ pc$^2$, i.e., through a 1.74–2.25 kpc radius. This represents a small fraction of the complete sampled area of 17,800 kpc$^2$, equivalent to a circle of radius 42 kpc. The emission is coming from $\sim$0.3% of the ALMA sampled area! Given that the optical size of the galaxy is even larger, the emission is coming from a tiny part compared to the optical emission from Malin 1.

It is useful to compare the detection limit of our observations and those obtained in 2022 with the 100 m Byrd Green Bank Telescope (GBT; G. Galaz et al. 2022). Although the comparison is not straightforward given the different structural principles between the GBT and ALMA (GBT a single dish and ALMA an interferometer), we can have a first approximation of the corresponding sensitivities. With the GBT, authors quote a sensitivity upper limit of $\sim$90 mK km s$^{-1}$ (at $1\sigma$ significance level), when considering a $7''$ spatial resolution and assuming a CO line width equal to the H I line width of 320 km s$^{-1}$ (F. Lelli et al. 2010). This corresponds to a GBT limit for the $^{12}$CO (1–0) line in the core of Malin 1 of $\sim$60 mK km s$^{-1}$ ($1\sigma$) with a line width of $\sim$140 km s$^{-1}$, which is equivalent to $\sim$40 mJy km s$^{-1}$ ($1\sigma$), comparable to our ALMA detection limit. However, the calibration uncertainty with the GBT at 110 GHz is of order 30%–50%, whereas the calibration of ALMA is much better defined, with figures of $\sim$10%.

### 3.2. Significance of Detections and Tentative Detections

What is presented above refers to the sources detected in the nuclear region of Malin 1 in both continuum and line emission as detections based on their high-S/N measurement. Here we provide more details about the molecular emission search process and about robust and tentative detections.

To search for blind detections of emission lines in the spectral cubes, we used LineSeeker[21] (J. González-López et al. 2019). Basically, it uses the negative side of spectral cubes as a reference for the noise and the number of expected false detections for a given emission-line S/N and width.

LineSeeker finds two high-significance emission-line detections with S/N values of 8.3 and 6.5. The former corresponds to a foreground galaxy with a velocity offset of $\approx -2500$ km s$^{-1}$ from the Malin 1 redshift (assuming that it corresponds to $^{12}$CO (1–0)), and the latter corresponds to the nuclear detection of $^{12}$CO (1–0) in Malin 1. The search hence returns two detections of emission lines in the spectral cube set

---

[20] https://science.nrao.edu/facilities/vla/proposing/TBconv
[21] https://github.com/jigonzal/LineSeeker

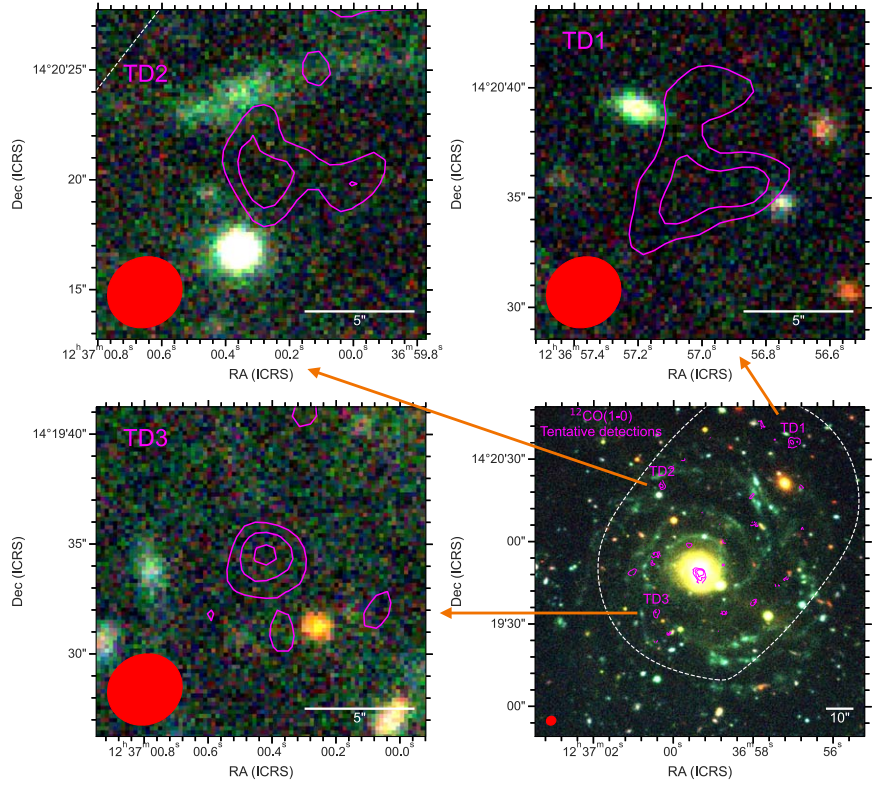

**Figure 3.** Disk location of the three tentative detections. We show here detections with $3 \leqslant S/N \leqslant 4$ of the peak of the collapsed lines, as indicated in the bottom right panel, where they are superimposed over a composite optical image (S. Boissier et al. 2016). Line detections denoted as TD1, TD2, and TD3 are amplified in the other panels. The corresponding line emission profiles are shown in Figure 4. Other fainter emissions in the disk have S/N smaller than 2 and are not considered, but they are also shown in this figure.

to detect $^{12}$CO (1–0) in Malin 1, and only one is associated with it.

A second search for fainter low-significance detection of emission lines was done by applying adaptive masking of a smaller spectral cube covering $\pm 1000$ km s$^{-1}$ around the central velocity of the nuclear $^{12}$CO (1–0) in Malin 1. We use the same procedure used by M. Solimano et al. (2024) to retrieve faint and extended emissions in ALMA observations. The method consists of applying a Gaussian convolution in both the spectral and spatial axes, measuring an S/N in the new convoluted cube, and then using all the voxels with S/N $\geqslant 3$ to create a moment 0. In this search we used a spectral Gaussian of $\sigma = 50$ km s$^{-1}$ and a spatial Gaussian of $\sigma = 1\rlap{.}''5$. The resulting moment-0 map is shown in Figure 3, where three (tentative) detections (TD1, TD2, and TD3) are marked. The corresponding line emission is shown in Figure 4. These have $3 \leqslant S/N \leqslant 4$, not high enough to claim a blind detection, and thus we identify them as "tentative detections."

Figure 3 shows that the three tentative detections do not have clear counterparts in the optical images and exhibit an offset from the main spiral arms of Malin 1. Although the lack of counterparts in the optical image would support the idea of the tentative detection being false, there are documented cases where significant offsets between CO and H$\alpha$ emission in particular have been observed in spiral galaxies (F. Egusa et al. 2004). In these cases, however, the shift in radial velocity relative to the central heliocentric radial velocity of the galaxy makes the subject a little bit more complex, and this will be discussed in Section 4.

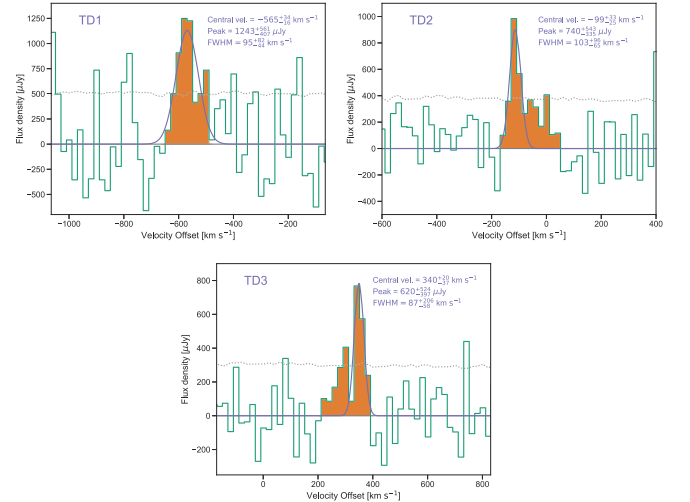

**Figure 4.** $^{12}$CO (1–0) line profiles of tentative detections. These correspond to the emissions shown in Figure 3. These detections have $3 \leqslant S/N \leqslant 4$. Some features of the line fitting are shown in the upper right corner of each line profile.

## 4. Discussion

In this section we discuss relevant quantities derived from the measured line fluxes, like the CO luminosity[22], the molecular gas mass, the molecular gas surface density, and so on. We shall also discuss the possible nature of tentative detections presented in Section 3.2.

---

[22] Note that CO luminosity is denoted as $L'$, and has units of kelvin kilometers per second per square parsec.

Table 1
Line Parameters for $^{12}$CO (1–0) Detections in the Center and Disk Regions for Malin 1

| ID | R.A. (2000.0) | Decl. (2000.0) | Central Vel. (km s$^{-1}$) | FWHM (km s$^{-1}$) | Flux (mJy km s$^{-1}$) | $L'_{CO}$ ($\times 10^7$ K km s$^{-1}$ pc$^{-2}$) | H$_2$ Mass ($\times 10^8\,M_\odot$) |
| --- | --- | --- | --- | --- | --- | --- | --- |
| (1) | (2) | (3) | (4) | (5) | (6) | (7) | (8) |
| Nucleus | 12:36:59 | +14:19:49 | $-140^{+22}_{-20}$ | $138^{+76}_{-43}$ | $142^{+21}_{-21}$ | $4.8^{+0.7}_{-0.7}$ | $2.1^{+0.3}_{-0.3}$ |
| TD1 | 12:36:57 | +14:20:36 | $-565^{+34}_{-16}$ | $95^{+82}_{-44}$ | $125^{+54}_{-38}$ | $4.1^{+1.8}_{-1.2}$ | $1.8^{+0.8}_{-0.5}$ |
| TD2 | 12:37:00 | +14:20:20 | $-99^{+33}_{-25}$ | $103^{+96}_{-65}$ | $75^{+35}_{-26}$ | $2.5^{+1.2}_{-0.8}$ | $1.1^{+0.5}_{-0.4}$ |
| TD3 | 12:37:00 | +14:19:34 | $340^{+20}_{-37}$ | $87^{+206}_{-58}$ | $56^{+25}_{-18}$ | $1.9^{+0.8}_{-0.6}$ | $0.8^{+0.4}_{-0.3}$ |

**Note.** Values are obtained using Gaussian fitting. See Sections 2 and 3 for details. Refer to Figures 1 and 3 for the corresponding identification region of emission in the galaxy. Column (1): source ID. Column (2): R.A. Column (3): decl. Column (4): heliocentric central velocity with respect to $z = 0.082702$. Column (5): FWHM of a Gaussian fit to the emission line. Column (6): integrated flux of the emission line in mJy km s$^{-1}$. Column (7): CO Luminosity. Column (8): molecular mass assuming $\alpha_{CO} = 4.4\,M_\odot$ (K km s$^{-1}$ pc$^2$)$^{-1}$.

Considering a sampled area of $\sim 19$ arcsec$^2$ in the central region ($\sim 47$ kpc$^2$), we obtain a CO luminosity $L'_{CO} = 4.8 \pm 0.7 \times 10^7$ K km s$^{-1}$ pc$^{-2}$ and a total mass for the molecular gas $M_{H_2} = (2.1 \pm 0.3) \times 10^8\,M_\odot \times (\alpha_{CO}/4.4)$. We use a conversion factor $\alpha_{CO}$ between the integrated $^{12}$CO (1–0) line intensity and the H$_2$ mass equal to $4.4\,M_\odot$ (K km s$^{-1}$ pc$^2$)$^{-1}$ (see Table 1 and A. D. Bolatto et al. 2013).

With the total area of the central region estimated to be within 47 kpc$^2$, we can compute the molecular gas mass surface density $\Sigma_{Mol}$. This returns $\Sigma_{Mol} \sim 5\,M_\odot$ pc$^{-2}$. This value can go down to $\Sigma_{Mol} \sim 1\,M_\odot$ pc$^{-2}$ if $\alpha_{CO} = 1\,M_\odot$ (K km$^{-1}$ pc$^{-2}$)$^{-1}$ is used. It is worth noting that our figures are in agreement with the corresponding value estimated by S. Boissier et al. (2016; $\Sigma_{H\,I+Mol} \sim 5\,M_\odot$ pc$^{-2}$) based on H I density and stellar evolution synthesis model fitting.

If we consider the total or global sampled area with ALMA (the dashed region in Figure 1, $\approx 17{,}800$ kpc$^2$), the molecular gas mass surface density $\Sigma_{Mol}$ decreases to $\sim 0.01\,M_\odot$ pc$^{-2} \times (\alpha_{CO}/4.4)$. This figure can be considered as a lower limit, since some emission could be still undetected, with line fluxes lower than our detection threshold. At face value, this value is roughly 10 times smaller than the upper limits computed by other authors (J. Braine et al. 2000; G. Galaz et al. 2022) and perhaps is one of the lowest global molecular gas surface densities computed to date from observations for a spiral galaxy.

It is also interesting to note that the SFR in outer spiral regions of Malin 1 as observed with the MUSE data in Junais et al. (2024) is much fainter than in the center. In fact, the H$\alpha$ SB is more than three times fainter in the arms than in the center, even for the brightest regions, and up to 100 times fainter for the faintest regions. In other words, if we naively assume that the CO SB is dimmed the same way as H$\alpha$ brightness, we should not expect to have an S/N larger than $\sim 2$ in CO for the brightest regions in the spiral arms, even if Malin 1 lies on the Kennicutt–Schmidt (K-S) law.

If we scale the mass of H I (T. E. Pickering et al. 1997; F. Lelli et al. 2010) to the same area where the CO is detected, we obtain a molecular-to-atomic gas mass ratio close to unity. However, if we consider the total sampled area, the molecular-to-atomic gas mass ratio $M_{H_2}/M_{H\,I}$ decreases to $\sim 0.01/2.3 = 0.004$, where $2.3\,M_\odot$ pc$^{-2}$ is the atomic gas mass surface density[23] ($\Sigma_{H\,I}$) for Malin 1 used by G. Galaz et al. (2022) and obtained from data of T. E. Pickering et al. (1997), which in turn are similar to those in F. Lelli et al. (2010). $M_{H_2}/M_{H\,I} = 0.004$ is an extremely low value, and it is a clear indication of the scarcity of the molecular gas in Malin 1 along its enormous extension.

The obtained molecular gas mass surface density $\Sigma_{Mol}$ can be used to better understand the star-forming properties of Malin 1. These properties can be compared to those computed for other galaxies. In particular, it is worth comparing the obtained values here with those computed in star-forming spiral galaxies where the molecular gas has been studied, along with the SFR surface density of these galaxies. Figure 5 shows the position of Malin 1 in the so-called K-S diagram (R. C. Kennicutt 1998; M. R. Krumholz et al. 2009), but in terms of the molecular gas surface density $\Sigma_{Mol}$ instead of the typical atomic gas surface density $\Sigma_{H\,I}$, compared to the values obtained for the sample of galaxies detected within the PHANGS-ALMA survey (J. Sun et al. 2023).

Considering the total surveyed area, it is clear that Malin 1 lies at the bottom end of the star-forming sequence but with very short molecular gas depletion times (green star in Figure 5). On the other hand, when only the central region is considered to compute the area where also the molecular gas surface density is much larger, the SFR surface density increases to the level of normal galaxies (magenta star in Figure 5). In other words, the central region of Malin 1 seems to convert molecular gas into stars not too differently from what is observed in normal spiral galaxies. From one side this supports what has been suggested earlier: the central part of Malin 1 behaves like an individual spiral galaxy, regardless of what is happening far outside in its diffuse disk (A. J. Barth 2007). For reference, recall that in optical bands the central part of Malin 1 has a similar size of the MW, i.e., $\simeq 30$ kpc. However, the fact that for more than 30 yr different authors and instruments have failed to detect CO tells us that, still, the SFR surface density in the very center of Malin 1 is not high ($\sim 10^{-3}\,M_\odot$ yr$^{-1}$ kpc$^{-2}$), as Figure 5 shows.

Given the outstanding spatial resolution of MUSE at the VLT and ALMA, allowing us to combine H$\alpha$ data from the first instrument with CO data from the second one, we are able to trace the SFR surface density as a function of the molecular gas density along different radii in the central part of Malin 1, as the colored line shows in Figure 5. We see that, in spite of the relatively narrow range of space within the center of Malin 1 ($\sim 20$ kpc), the SFR surface density drops from $\sim 2 \times 10^{-3}\,M_\odot$ yr$^{-1}$ kpc$^{-2}$ to $\sim 4 \times 10^{-5}\,M_\odot$ yr$^{-1}$ kpc$^{-2}$, with molecular gas mass densities varying from $\sim 2$ to $\sim 0.8\, M_\odot\, {\rm pc}^{-2}$, respectively.

---

[23] We use here the same cosmology as in G. Galaz et al. (2022).

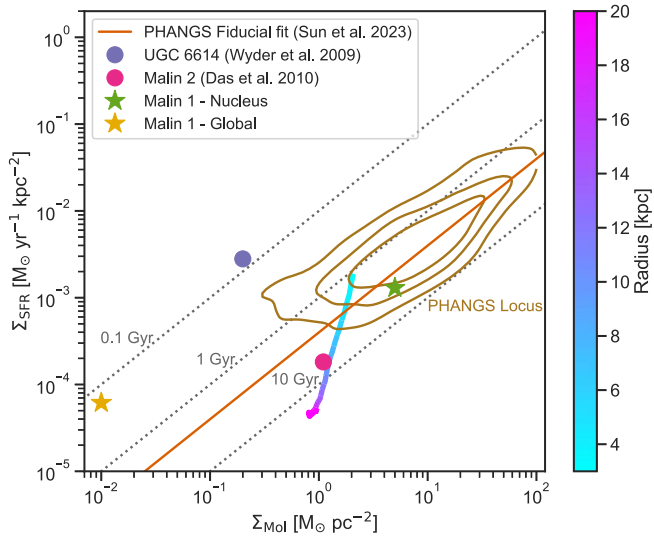

**Figure 5.** K-S diagram for the molecular gas in Malin 1 (R. C. Kennicutt 1998; M. R. Krumholz et al. 2009). This plot shows the range of stellar formation surface density as a function of the molecular gas surface density for different area coverage and detections. The star formation surface density $\Sigma_{\rm SFR}$ in the very center (magenta star) is $\sim 2 \times 10^{-3}\, M_\odot\, {\rm yr}^{-1}\, {\rm kpc}^{-2}$ and has a molecular gas mass surface density $\Sigma_{\rm Mol} \sim 5\, M_\odot\, {\rm pc}^{-2}$, compared to the global $\Sigma_{\rm SFR}$ (i.e., considering the total surveyed area; green star) of $\sim 6 \times 10^{-5}\, M_\odot\, {\rm yr}^{-1}\, {\rm kpc}^{-2}$ with a corresponding molecular gas surface density of $\sim 0.01\, M_\odot\, {\rm pc}^{-2}$. The thin dotted lines represent linear relations with constant molecular gas depletion times of 0.1, 1, and 10 Gyr. The density contours (40%–80%–95% levels from inside to outside contours) represent the distributions of all 1.5 kpc scale regions with $>3\sigma$ detections for both $\Sigma_{\rm Mol}$ and $\Sigma_{\rm SFR}$ for galaxies studied in the PHANGS-ALMA sample (J. Sun et al. 2023). The colored-gradient trace represents the $5\sigma$ upper limit molecular gas surface density combined with the corresponding SFR surface density measured with MUSE (Junais et al. 2024). The different radii for the estimates are shown in the color bar on the right. We also show the position of the gLSBGs UGC 6614 and Malin 2 (M. Das et al. 2006, 2010; T. K. Wyder et al. 2009). See Section 4 for further details.

In Figure 5 we also show the position of gLSBGs UGC 6614 and Malin 2, two of the few gLSBGs with measured CO emission (M. Das et al. 2006, 2010; T. K. Wyder et al. 2009). For UGC 6614, $\Sigma_{\rm H_2}$ represents the global molecular gas mass surface density, i.e., the quantity obtained considering the total surveyed area. For Malin 2, we plot the molecular gas mass surface density obtained considering only the region where the emission is detected (as in our case for Malin 1 with the colored bar). For the global emission, UGC 6614 still has a global $\Sigma_{\rm Mol} \sim 0.2\, M_\odot\, {\rm pc}^{-2}$, a value 20 times larger than the one we obtain here for Malin 1 (green star). The case of Malin 2 seems not too different from that for Malin 1, although Malin 2 is half the size of Malin 1.

Now we discuss the possible nature of the $^{12}$CO (1–0) emission from regions TD1, TD2, and TD3 (what we dubbed "tentative detections"). Figure 4 shows the corresponding line profiles for our tentative detections, which show in addition for TD1 and TD3 large radial velocity shifts relative to the systemic velocity of Malin 1 (see also Table 1, which presents some properties of the $^{12}$CO (1–0) line detections). F. Lelli et al. (2010) improved the H I velocity field from T. E. Pickering et al. (1997) and concluded that there is up to $\pm 200\, {\rm km\, s}^{-1}$ difference between the systemic velocity of Malin 1 and the northern (positive velocities) and southern (negative velocities) extreme sides of the disk in Malin 1, compatible with the observations performed in ionized gas with MUSE (E. J. Johnston et al. 2024), which also confirms the result of Junais et al. (2020) that high velocities are reached close to the center. This suggests that TD1 and TD3 are exceeding the H I radial velocity by more than $\sim 400\, {\rm km\, s}^{-1}$, making it therefore difficult to associate these two emissions with Malin 1, and making these emissions uncertain from the point of view of their actual nature. Note that T. E. Pickering et al. (1997) and F. Lelli et al. (2010) also identify a potential H I warp at those distances (PA variations), further revealing the complexity of this large disk structure.

The case of TD2 is more favorable to belong to the Malin 1 disk, especially when comparing the velocity difference with the central detection of $\sim +40\, {\rm km\, s}^{-1}$ with the expected velocity shift that is close to $+100\, {\rm km\, s}^{-1}$ in that region. In any case, the question still stands: if TD1 and TD3 emissions do not come from Malin 1, where do they come from? Although this issue escapes the scope of this Letter, we include these results here. An element that could shed some light on this issue is the possible association of these emissions with high-velocity clouds in Malin 1. Such clouds are not necessarily associated with optical counterparts and have been observed in the Local Group and in the MW, and they are the subject of discussions (J. N. Bregman 1980; B. P. Wakker & H. van Woerden 1997; L. Blitz et al. 1999). Furthermore, our preliminary semianalytic orbital calculations (R. Bustos-Espinoza et al. 2025, in preparation) suggest that TD2 and TD3 could have orbits bound to the Malin 1 system, while TD1 is probably associated with an external source.

An interesting possibility for Malin 1 is the presence of CO-dark molecular gas (DMG), which refers to molecular gas not traced by CO emission. This gas has been identified in the solar neighborhood and recently in the MW, 13 kpc from the Galactic center (G. Luo et al. 2024). This raises the potential that Malin 1 is replenished with DMG, making CO a poor tracer. Such a scenario is plausible given that the visual extinction, H$_2$ density, and molecular fraction of DMG are similar to those in nearby diffuse molecular clouds. L. Ramambason et al. (2024) also explore stellar formation in low-metallicity galaxies and highlight the role of DMG. They conclude that low-metallicity galaxies with high clumpiness parameters may exhibit $\alpha_{\rm CO}$ values comparable to the Galactic value, even at low metallicity. While Malin 1 is not significantly low in metallicity, it may possess a turbulent interstellar medium (ISM; indicated by velocity dispersions) and a high conversion factor, potentially eliminating the need to invoke DMG. However, confirming clumpiness requires higher-resolution observations.

This specific topic, together with other secure detections associated with other sources in the field, will be investigated in further detail in a forthcoming paper.

## 5. Summary and Conclusions

In this work we have finally detected CO in the iconic gLSBG Malin 1. We detected continuum and $^{12}$CO (1–0) emissions in the center of Malin 1, in a position that is consistent with findings and analysis of H$\alpha$ emission recently obtained (E. J. Johnston et al. 2024; Junais et al. 2024). The total central continuum flux density is $107 \pm 12\, \mu{\rm Jy}$, and the velocity-integrated flux density for the $^{12}$CO (1–0) line is $143 \pm 41\, {\rm mJy\, km\, s}^{-1}$. This yields a CO luminosity of $L'_{\rm CO} = (4.8 \pm 0.7) \times 10^7\, {\rm K\, km\, s}^{-1}\, {\rm pc}^{-2}$. The total estimated molecular mass is $\sim (2.1 \pm 0.3) \times 10^8\, M_\odot \times (\alpha_{\rm CO}/4.4)$, and the central molecular gas mass surface density is $\Sigma_{\text{Mol}} \sim 3-5\ M_\odot\ \text{pc}^{-2}$. However, when considering the total sampled area, $\Sigma_{\text{Mol}}$ is reduced to $\sim 0.01\ M_\odot\ \text{pc}^{-2}$. With these figures, the molecular-to-atomic gas mass ratio is $\sim 0.04$ for Malin 1.

We also report on three tentative detections in the disk. Because of large velocity shifts relative to the systemic velocity of Malin 1, only one of these detections seems to be robust and can be associated with Malin 1.

In this work we provide solid evidence of the existence of molecular gas in the largest spiral galaxy observed so far. The goal to detect CO in Malin 1 has now been reached. Results presented here show not only that Malin 1 has a very low global SFR surface density and extremely low global molecular gas surface density, as former upper limits anticipated (J. Braine et al. 2000; G. Galaz et al. 2022), but also that the stellar formation is sparse and at the same time spotty, along its vast physical extension. This is consistent with conclusions of other authors (J. Schombert & S. McGaugh 2014; S. Boissier et al. 2016) who claim an intermittent star formation history, supported by colors of individual regions. Nevertheless, our results, along with those obtained recently (Junais et al. 2024), suggest that galaxies as large and diffuse as Malin 1 are forming stars in a significant number of spots along their huge disks, in spite of their extremely low global molecular gas surface densities compared to other galaxies. The emerging picture is that these star-forming spots seem quite isolated, with star-forming processes that seem to challenge typical scales observed in HSB spiral galaxies. A more profound understanding of the actual conversion of molecular gas into stars taking place at these gas densities requires more precise and higher-resolution millimeter observations. This includes an actual determination of the conversion factor at these extremely low molecular gas mass surface densities and likely different ISM temperatures. The complete physical processes that explain the distribution of the molecular gas in Malin 1 are still unclear, but should be better understood in future works combining higher resolution observations and refined modeling.

## Acknowledgments

We thank the anonymous referee for their insightful suggestions, which improved the quality and clarity of this Letter. We also appreciate the editor's willingness for the prompt handling of this Letter and the help from the ApJ Letters staff. G.G., V.G., T.P., R.B., K.J., and E.J.J. gratefully acknowledge support by the ANID BASAL projects ACE210002 and FB210003. G.G. acknowledges the support of Pontificia Universidad Católica de Chile through a sabbatical grant, as well as the European Southern Observatory (ESO) for support during the completion of this work. E.J.J. acknowledges support from FONDECYT Iniciación en Investigación 2020 Project 11200263. J. is funded by the European Union (MSCA EDUCADO, GA 101119830 and WIDERA ExGal-Twin, GA 101158446). This Letter makes use of the following ALMA data: ADS/JAO. ALMA#2023.1.01105.S. ALMA is a partnership of ESO (representing its member states), NSF (USA) and NINS (Japan), together with NRC (Canada), NSTC and ASIAA (Taiwan), and KASI (Republic of Korea), in cooperation with the Republic of Chile. The Joint ALMA Observatory (JAO) is operated by ESO, AUI/NRAO and NAOJ. The National Radio Astronomy Observatory (NRAO) is a facility of the National Science Foundation operated under cooperative agreement by Associated Universities, Inc. This research has made use of the Astrophysics Data System, funded by NASA under Cooperative Agreement 80NSSC21M00561. This research has made use of the Sloan Digital Sky Survey (SDSS) data. Funding for the SDSS has been provided by the Alfred P. Sloan Foundation, the Participating Institutions, the National Science Foundation, the US Department of Energy, the National Aeronautics and Space Administration, the Japanese Monbukagakusho, the Max Planck Society, and the Higher Education Funding Council for England. The SDSS website is http://www.sdss.org/. The SDSS is managed by the Astrophysical Research Consortium for the Participating Institutions.

### ORCID iDs

Gaspar Galaz ⬤ https://orcid.org/0000-0002-8835-0739
Jorge González-López ⬤ https://orcid.org/0000-0003-3926-1411
Viviana Guzmán ⬤ https://orcid.org/0000-0003-4784-3040
Hugo Messias ⬤ https://orcid.org/0000-0002-2985-7994
Junais ⬤ https://orcid.org/0000-0002-7016-4532
Samuel Boissier ⬤ https://orcid.org/0000-0002-9091-2366
Benoît Epinat ⬤ https://orcid.org/0000-0002-2470-5756
Peter M. Weilbacher ⬤ https://orcid.org/0000-0003-4766-902X
Thomas Puzia ⬤ https://orcid.org/0000-0003-0350-7061
Evelyn J. Johnston ⬤ https://orcid.org/0000-0002-2368-6469
Philippe Amram ⬤ https://orcid.org/0000-0001-5657-4837
David Frayer ⬤ https://orcid.org/0000-0003-1924-1122
Matìas Blaña ⬤ https://orcid.org/0000-0003-2139-0944
J. Christopher Howk ⬤ https://orcid.org/0000-0002-2591-3792
Michelle Berg ⬤ https://orcid.org/0000-0002-8518-6638
Roy Bustos-Espinoza ⬤ https://orcid.org/0000-0002-6208-3337
Paulo Cortés ⬤ https://orcid.org/0000-0002-3583-780X
Diego García-Appadoo ⬤ https://orcid.org/0000-0003-1595-8548
Katerine Joachimi ⬤ https://orcid.org/0000-0001-5348-7543